\documentclass{article}
\usepackage{epsfig}
\usepackage{amsmath}
\usepackage{amsfonts}\tolerance=10000
\date{\today} 
 
\usepackage{graphicx}

\newcommand{\insertplot}[5]{\begin{figure}
 \hfill\hbox to 0.05in{\vbox to #5in{\vfill
 \inputplot{#1}{#4}{#5}}\hfill}
 \hfill\vspace{-.1in}
 \caption{#2}\label{#3}
 \end{figure}}
 \newcommand{\inputplot}[3]{
 \special{ps: plotfile #1}
}
\newcounter{fig}

\newcommand{\beq}{\begin{equation}}
\newcommand{\eeq}{\end{equation}}
\newcommand{\beqs}{\begin{eqnarray}}
\newcommand{\eeqs}{\end{eqnarray}}

\newcommand{\ra}{\rightarrow}
\numberwithin{equation}{section}
\newcommand{\be}{\begin{equation}}
\newcommand{\ee}{\end{equation}}
\newcommand{\bea}{\begin{eqnarray}}
\newcommand{\eea}{\end{eqnarray}}

\usepackage{graphicx}

\begin{document}

\title{Five-dimensional rotating black holes in \\
Einstein-Gauss-Bonnet theory } 
 
\author{{\large Yves Brihaye}$^{\dagger}$
and {\large Eugen Radu}$^{\ddagger }$ \\ \\
$^{\dagger}${\small Physique-Math\'ematique, Universite de
Mons-Hainaut, Mons, Belgium}\\
$^{\ddagger}${\small Laboratoire de Math\'ematiques et Physique Th\'eorique,
Universit\'e Fran\c{c}ois-Rabelais, Tours, France}
  }

\maketitle 
 
\begin{abstract}
We present arguments for the existence of  five-dimensional
rotating black holes 
with equal magnitude angular momenta in Einstein-Gauss-Bonnet theory with negative cosmological
constant.
These solutions posses
a regular horizon of spherical topology and 
approach asymptotically an Anti-de Sitter spacetime background. 
We discuss the general properties of these solutions and, using an adapted counterterm prescription, 
we compute their entropy and
conserved charges.
\end{abstract}

\section{Introduction}

In five dimensions, the most general theory of gravity leading to
second order field equations for the metric is the so called 
Einstein-Gauss-Bonnet (EGB) theory, which contains quadratic powers of the 
curvature.
The  Gauss-Bonnet term appears as the first curvature stringy
correction to general relativity~\cite{1,Myers:1987yn}, when assuming
that the tension of a string is
large as compared to the energy scale of other variables.

The study of black holes with higher derivative curvature
in Anti-de Sitter (AdS) spaces has been considered by many authors
 in the recent years.
Static AdS black hole solutions in EGB gravity 
are known  in closed form, presenting
a number of interesting features (see $e.g.$
\cite{Deser}, \cite{wheeler}, \cite{neupane}, \cite{Torii:2005xu}
and the references therein).
It is of interest to generalize these solutions by including the effects of rotation.
This problem has been considered recently in \cite{Kim:2007iw} within a perturbative
approach.
The authors of \cite{Kim:2007iw} discussed some properties of a particular set of 
asymptotically AdS$_d$ $(d>4)$ rotating solutions in EGB theory with one nonvanishing angular momentum 
(where the rotation parameter appears as a small quantity), 
the effects of an U(1) field being also included.

The main purpose of this work
is to  present  numerical
evidence for the existence of a different class
of rotating solutions in $d=4+1$  EGB theory with negative cosmological constant,
approaching asymptotically an AdS spacetime background. 
These solutions are found within a nonperturbative approach, by directly solving the EGB equations
with suitable boundary conditions.
They posses a regular horizon of spherical topology  
and have two equal magnitude angular momenta.
This leads to a system of coupled nonlinear ordinary differential
equations (ODEs), which are solved numerically.
The same approach has been  employed recently to  construct
Einstein-Maxwell rotating black hole solutions in higher dimensions
 \cite{Kunz:2005nm,Kunz:2006eh}.

\section{The general formalism}
\subsection{The action and boundary counterterms }
We consider the EGB action with a negative cosmological constant $\Lambda=-6/\ell^2$ 
\begin{eqnarray}
\label{action}
I=\frac{1}{16 \pi G}\int_\mathcal{M}~d^5x \sqrt{-g} \left(R-2 \Lambda+\frac{\alpha}{4}L_{GB}\right),
\end{eqnarray}
where $R$ is the Ricci scalar and 
\begin{eqnarray}
\label{LGB}
L_{GB}=R^2-4R_{\mu \nu}R^{\mu \nu}+R_{\mu \nu \sigma \tau}R^{\mu \nu \sigma \tau},
\end{eqnarray}
is the Gauss-Bonnet term.
The variation of the action (\ref{action}) with respect to the metric
tensor results in the equations of the model
\begin{eqnarray}
\label{eqs}
R_{\mu \nu } -\frac{1}{2}Rg_{\mu \nu}+\Lambda g_{\mu \nu }+\frac{\alpha}{4}H_{\mu \nu}=0~,
\end{eqnarray}
where
\begin{equation}
\label{eq1}
H_{\mu \nu}=2(R_{\mu \sigma \kappa \tau }R_{\nu }^{\phantom{\nu}%
\sigma \kappa \tau }-2R_{\mu \rho \nu \sigma }R^{\rho \sigma }-2R_{\mu
\sigma }R_{\phantom{\sigma}\nu }^{\sigma }+RR_{\mu \nu })-\frac{1}{2}%
L_{GB}g_{\mu \nu }  ~.
\end{equation}
For a well-defined variational principle, one has to supplement the 
action (\ref{action}) with the Gibbons-Hawking surface term 
\begin{equation}
I_{b}^{(E)}=-\frac{1}{8\pi G}\int_{\partial \mathcal{M}}d^{4}x\sqrt{-\gamma }K~,
\label{Ib1}
\end{equation}
and its counterpart for the Gauss-Bonnet gravity  \cite{Myers:1987yn} 
\begin{equation}
I_{b}^{(GB)}=-\frac{\alpha}{16\pi G}\int_{\partial \mathcal{M}}d^{4}x\sqrt{-\gamma }%
 \left( J-2{\rm G}_{ab} K^{ab}\right)  ~,
\label{Ib2}
\end{equation}
where $\gamma _{ab }$ is the induced metric on the boundary,  $K$ is  the
trace of the extrinsic curvature of the boundary,
  ${\rm G}_{ab}$ is the Einstein tensor of the metric $\gamma _{ab}$ and $J$ is the
trace of the tensor
\begin{equation}
J_{ab}=\frac{1}{3}%
(2KK_{ac}K_{b}^{c}+K_{cd}K^{cd}K_{ab}-2K_{ac}K^{cd}K_{db}-K^{2}K_{ab})~.
\label{Jab}
\end{equation}

To compute the conserved charges of the asymptotically AdS
solutions in EGB gravity, we use the approach proposed by  Balasubramanian
and Kraus in \cite{Balasubramanian:1999re}.
This technique was inspired by AdS/CFT correspondence and consists 
in adding suitable counterterms $I_{ct}$
to the action of the theory in order to ensure the finiteness of the boundary
stress tensor 
$T_{ab}= \frac{2}{\sqrt{-\gamma}} \frac{\delta I}{ \delta \gamma^{ab}}$
derived by the quasilocal energy definition
\cite{Brown:1993br}. 

Therefore we supplement the general action (which contains the surface terms for
Einstein and Gauss-Bonnet gravity)  with  the following boundary counterterm 
\begin{eqnarray}
I_{\mathrm{ct}} &=&\frac{1}{8\pi G}\int_{\partial \mathcal{M}} d^{4}x\sqrt{-\gamma }
(
c_1-\frac{c_2}{2} \mathsf{R}
)~,
\end{eqnarray}
where $\mathsf{R}$ is the curvature scalar associated with the induced metric $\gamma $ 
(see also \cite{nojiri} for previous work on boundary conterterm technique in EGB gravity, 
applied to non-rotating solutions).
The consistency of the procedure fixes the expression\footnote{ As $\alpha\to 0$,
one recovers the known expression in Einstein gravity, $c_1\to -3/\ell+\alpha/4\ell^3+O(\alpha)^2,$
$c_2 \to \ell/2+3\alpha/8\ell+O(\alpha)^2.$}
of $c_1,c_2$:
\begin{eqnarray}
c_1=\frac{-1 - \frac{2\,\alpha }{\ell^2} + {\sqrt{1 - \frac{2\,\alpha }{\ell^2}}}}
  {{\sqrt{\alpha }}\,{\sqrt{1 - {\sqrt{1 - \frac{2\,\alpha }{\ell^2}}}}}},~
~~
c_2=\frac{{\sqrt{\alpha }}\,\left( 3 - \frac{2\,\alpha }{\ell^2} 
- 3\,{\sqrt{1 - \frac{2\,\alpha }{\ell^2}}} 
\right) }
  {2\,{\left( 1 - {\sqrt{1 - \frac{2\,\alpha }{\ell^2}}} \right) }^{\frac{3}{2}}}~.
\end{eqnarray}
Varying the total action  
with respect to the
boundary metric $\gamma_{ab}$, we find the 
divergence-free boundary stress-tensor
\begin{eqnarray}
T_{ab}=\frac{1}{8 \pi G}
\left(
K_{ab}-K\gamma_{ab}
+c_1\gamma_{ab}+c_2{\rm G}_{ab}
+ \frac{{\alpha}}{2} (Q_{ab}-\frac{1}{3}Q\gamma_{ab}) 
\right)~,
\end{eqnarray} 
where \cite{Davis:2002gn}
\begin{eqnarray}
Q_{ab}= 
2KK_{ac}K^c_b-2 K_{ac}K^{cd}K_{db}+K_{ab}(K_{cd}K^{cd}-K^2)
+2K \mathsf{R}_{ab}+\mathsf{R}K_{ab}
-2K^{cd}\mathsf{ R}_{cadb}-4 \mathsf{R}_{ac}K^c_b~,
\end{eqnarray}
with $\mathsf{R}_{abcd}$, $\mathsf{R}_{ab}$  denoting
the Riemann   and Ricci tensors of the boundary metric.
 
Provided the boundary geometry has an isometry generated by a
Killing vector $\xi ^{i}$, a conserved charge
\begin{eqnarray}
{\frak Q}_{\xi }=\oint_{\Sigma }d^{3}S^{i}~\xi^{j}T_{ij}
\label{charge}
\end{eqnarray} 
can be associated with a closed surface $\Sigma $ \cite{Balasubramanian:1999re}. 
Physically, this means that a collection of observers on
the hypersurface whose metric is $\gamma$ all observe the same value
of ${\frak Q}_{\xi }$ provided this surface has an isometry
generated by $\xi$. 

\subsection{The metric ansatz and known limits}
While the
general EGB rotating black holes would possess two independent
angular momenta and a more general  topology
of the event horizon, 
we  restrict here to configurations with two 
equal magnitude angular momenta and a spherical horizon topology.
The suitable metric ansatz\footnote{ 
EGB  rotating  topological black hole 
 with zero scalar curvature  of the
event horizon are known in closed
form (see $e.g.$ \cite{cai}, \cite{Dehghani:2006dh} and references there).
However, they are found for a different metric ansatz and present rather
different properties.} 
 reads  \cite{Kunz:2005nm}
\begin{eqnarray}
\label{metric}
&&ds^2 = \frac{dr^2}{f(r)}
  + g(r) d\theta^2
+h(r)\sin^2\theta \left( d \varphi -w(r)dt \right)^2 
+h(r)\cos^2\theta \left( d \psi -w(r)dt \right)^2 
\\
\nonumber
&&{~~~~~~}+(g(r)-h(r))\sin^2\theta \cos^2\theta(d \varphi -d \psi)^2
-b(r) dt^2
\end{eqnarray}
where $\theta  \in [0,\pi/2]$, $(\varphi,\psi) \in [0,2\pi]$, $r$ and $t$ being the
radial and time coordinates. 
For such solutions, the isometry group is enhanced from $R \times U(1)^{2}$
to $R \times U(2)$, where $R$ denotes the time translation.
This symmetry enhancement allows us to deal with  ODEs (in what follows, we fix the metric gauge
by taking $g(r)=r^2$).

For the metric ansatz (\ref{metric}), the EGB field equations (\ref{eqs})  
present   two well known exact 
solutions.
The first one corresponds to the generalization \cite{Deser} of the static 
Schwarzschild-AdS solution with a Gauss-Bonnet term\footnote{Note 
that the EGB gravity presents two kind of static black hole solutions, which are 
classified into the plus and the minus branches,
$
f_{(\pm)}(r)= b_{(\pm)}(r)=1 + \frac{r^2}{\alpha}
 \bigg(
1 \pm {\sqrt{1 + 2\alpha\,\left(  \frac{m}{r^4} - \frac{1}{\ell^{2}} \right) \, }}
 \bigg ).
$
 In this paper we shall restrict to the minus branch solutions, which present a well defined Einstein gravity limit.
}
\begin{eqnarray}
\label{SGB-AdS} 
f(r)= b(r)=1 + \frac{r^2}{\alpha}
\left(
1 - {\sqrt{1 + 2\alpha\,\left(  \frac{m}{r^4} - \frac{1}{\ell^{2}} \right) \, }}
\right ),~~~
g(r)=h(r)=r^2,~~w(r)=0~.
\end{eqnarray}
The AdS$_5$ generalization \cite{taylor:1999,Gibbons:2004js}
of the  Myers-Perry rotating
 black holes \cite{Myers:1986un}  with equal magnitude angular momenta
is found for $\alpha=0$ (no Gauss-Bonnet term) 	and has
\begin{eqnarray}
\label{MP-AdS}
f(r)=1
+\frac{r^2}{\ell^2}
-\frac{2{\hat M}\Xi}{r^{2}}
+\frac{2{\hat M}{\hat a}^2}{r^{4}},~
h(r)=r^2\left(1+\frac{2{\hat M}{\hat a}^2}{r^{4}}\right),~
w(r)=\frac{2{\hat M}{\hat a}}{r^2 h(r)},~
g(r)=r^2,~b(r)=\frac{r^2 f(r)}{h(r)},
\end{eqnarray}
where ${\hat M}$ and ${\hat a}$ are two constants related to the solution's mass-energy and 
angular momentum, while $\Xi=1-{\hat a}^2/\ell^2$.

\section{Black hole properties}
We are interested in black hole solutions with an
horizon located at a constant value of
the radial coordinate
$r=r_h>0$. Restricting  to
nonextremal solutions, the following expansion holds near the event horizon:
\begin{eqnarray}
\label{c1}
&f(r)=f_1(r-r_h)+  O(r-r_h)^2,h(r)=h_h+ O(r-r_h),
b(r)=b_1(r-r_h)+O(r-r_h)^2,w(r)=w_h+ O(r-r_h).~{~ }
\end{eqnarray}
The event horizon parameters $r_h,~f_1,~b_1$,~$w_h$ and $h_h$ 
(with $(f_1,~b_1,~h_h)>0$) are related in a complicated way to the global
charges of the solutions.
 
The metric functions have the following asymptotic behaviour
in 
terms of the constants\footnote{The MPAdS$_5$ solution (\ref{MP-AdS}) has $f_2=2{\hat M}({\hat a^2}/\ell^2-1)$,
$b_2=-2 {\hat M}$,
$w_4=2{\hat M} {\hat a}$.
For the static Schwarzschild-AdS-Gauss-Bonnet solution (\ref{SGB-AdS}) 
one finds $f_2=b_2=-m/\sqrt{1 - \frac{2\,\alpha}{\ell^2} },~w_4=0$.
} 
$f_2,~b_2 $ and $w_4$:
\begin{eqnarray}
\label{inf1}
&&f= 1+ \frac{r^2}{\alpha} \big({1-\sqrt{1 - \frac{2\,\alpha}{\ell^2} }}~\big)
+\frac{f_2}{r^2}+O(1/r^4),~~~
b= 1+ r^2 \frac{1-\sqrt{1 - \frac{2\,\alpha}{\ell^2} }}{\alpha}+\frac{b_2}{r^2}+O(1/r^4),
\\
\nonumber
&&h= r^2+\ell^2\frac{f_2-b_2}{2r^2}\left( 1 + {\sqrt{1 - \frac{2\,\alpha}{\ell^2} }} ~\right)  +O(1/r^6),
~~~
w(r)=\frac{w_4}{r^4}+O(1/r^8)~.
\label{exp_inf}
\end{eqnarray}
One can see that, similar to the static case, the parameter $\alpha$ must obey $\alpha\leq \ell^2/2$,
beyond which the theory is undefined.
For these asymptotics, the effective cosmological constant is $\Lambda_{eff}=\Lambda(1+\sqrt{1 - \frac{2\,\alpha}{\ell^2} })/2.$ 

The Killing vector  $\chi=\partial/\partial_t+
 \Omega_\varphi \partial/\partial \varphi + \Omega_\psi \partial/\partial \psi $ is 
orthogonal to and null on the horizon. For the solutions 
within the ansatz (\ref{metric}), the 
event horizon's angular velocities are 
all equal, $\Omega_\psi=\Omega_\varphi=\omega_h$.
The Hawking temperature as  found by computing
the surface gravity is 
\begin{eqnarray} 
\label{Temp-rot} 
  T_H=\frac{\sqrt{b_1f_1}}{4\pi}.
\end{eqnarray} 
Another quantity of interest is
the area $A_H$ of the rotating black hole horizon 
\begin{eqnarray}
\label{A2} 
A_H= \sqrt{h_h}  r_h^2 V_3,
\end{eqnarray} 
where $V_3=2\pi^2$ denotes the area of the unit three dimensional sphere.

These rotating solutions present also an ergoregion
inside of which the observers 
cannot remain stationary, and will move in the direction of rotation.
The ergoregion is the region bounded by the event horizon, located 
at $r=r_h$ and the stationary limit surface, 
or the ergosurface.
 The Killing 
vector $\partial/\partial t$ becomes null on the  ergosurface , $i.e.$ 
$g_{tt}= -b(r)+h(r) w(r)^2=0$. 
The  ergosurface does not interesect the horizon. %

\subsection{The global charges and entropy of solutions}
The global charges of these solutions are computed by using the
counterterm formalism\footnote{A different computation of the mass and angular momentum of EGB solutions was also reported namely in
\cite{deruelle:2005}, \cite{tekin}, \cite{kofinas}.} 
presented in Section 2.
The computation of the boundary stress-tensor
$T_{ab}$ is straightforward and we find the 
following expression for the  components of interest here
\begin{eqnarray}
\label{tik} 
 &&T_\varphi^t=\frac{1}{8\pi G}\sqrt{\frac{\alpha(1-\frac{2\alpha}{\ell^2})}{1-\sqrt{1-\frac{2\alpha}{\ell^2}}}}
 \frac{2w_4\sin^2 \theta}{r^4}+O(1/r^6),~~
  T_\psi^t=\frac{1}{8\pi G}\sqrt{\frac{\alpha(1-\frac{2\alpha}{\ell^2})}{1-\sqrt{1-\frac{2\alpha}{\ell^2}}}}
 \frac{2w_4\cos^2 \theta}{r^4}+O(1/r^6),
 \\
 \nonumber
 &&T_t^t=-\frac{1}{8\pi G}\frac{\sqrt{\alpha}}{8(1-\sqrt{1-\frac{2\alpha}{\ell^2}}~)^{3/2}}
 \left(3\alpha(-2+3\sqrt{1-\frac{2\alpha}{\ell^2}}~) 
 +4\big(-1+\frac{2\alpha}{\ell^2}+\sqrt{1-\frac{2\alpha}{\ell^2}}~\big)(f_2-4b_2)\right)
 \frac{1}{r^4}+O(1/r^6).
\end{eqnarray}
The mass-energy $E$ of solutions is the 
charge associated with the Killing vector $\partial/\partial t$,
\begin{eqnarray}
\label{mass}
&&E=E^{(0)}+E^{(c)},~~{\rm where }~~
\\
\nonumber
&&E^{(0)}=\frac{V_3}{8 \pi G}\frac{(f_2-4b_2)}{2 }{\sqrt{1 - \frac{2\,\alpha}{\ell^2} }}~,
~~~
 E^{(c)}=\frac{V_3}{8 \pi G}
\frac{3\ell^2}{16}
\left(
 1- \frac{6\,\alpha}{\ell^2}+ {\sqrt{1 - \frac{2\,\alpha}{\ell^2} }}  
\right)~,
\end{eqnarray} 
where $ E^{(c)}$ represents the Casimir term \cite{Balasubramanian:1999re} in EGB gravity, presenting a nontrivial
$\alpha$ dependence\footnote{In the small $\alpha$ limit, one finds $ E^{(c)}=3\pi\ell^2/32 G- 21\alpha \pi/64 G+O(\alpha)^2.$
Note that the first order correction to the mass of
pure global AdS$_5$ does not depend on the value of the cosmological constant.}
(this term appears also in the static limit (\ref{SGB-AdS})).
These black holes have also two equal magnitude  angular momenta $J_\varphi=J_\psi=J$, with
\begin{eqnarray}
\label{Ji} 
J=\frac{V_3}{8 \pi G}  w_4\sqrt{1 - \frac{2\,\alpha}{\ell^2} }~,
\end{eqnarray}
representing the 
charges
associated with the Killing vectors $\partial/\partial \varphi$, $\partial/\partial \psi$ as
computed according to (\ref{charge}). 
 
 Also, in what follows it is important to use the observation that one can write
\begin{eqnarray}
\label{totder1} 
&\frac{1}{\sin^2 \theta}(R_\varphi^t+\frac{\alpha}{4}H_\varphi^t)=
\frac{1}{\cos^2 \theta}(R_\psi^t+\frac{\alpha}{4}H_\psi^t)=  \frac{1}{2 r^2} \sqrt{\frac{f}{ b h}}\frac{d }{dr}  
 \left (\sqrt{\frac{f h}{b}} h w'
 \bigg(
- r^2 +  \alpha ( f-4+ \frac{3h}{r^2})\bigg)
\right ),
\end{eqnarray} 
\begin{eqnarray}
\label{totder2} 
\nonumber
R_t^t+\frac{\alpha}{4}(H_t^t+\frac{1}{2}L_{GB})=
\frac{1}{2r^2} \sqrt{\frac{f}{ b h}}  \frac{d }{dr}
 \left  (\sqrt{\frac{f h}{b}}
 \bigg[r^2(hww'-b')+\alpha\left((f-4+\frac{h}{r^2}+\frac{rfh'}{h})b'
 +(4-\frac{3h}{r^2}-f)hww'+rfhw'^2\right)
  \bigg]
\right ) .
\end{eqnarray} 
The
gravitational thermodynamics of the EGB black holes can be  formulated via the 
path integral approach \cite{GibbonsHawking1,Hawking:ig}. 
However, while the static vacuum Lorentzian 
solutions (\ref{SGB-AdS}) extremize also the Euclidean
 action as the analytic continuation in time has no effect at the level
 of the equations of motion, this is not the case of 
the rotating configurations discussed in this paper. In this case it 
is not possible to find directly real solutions on the 
Euclidean section by Wick rotating $t\ra i\tau$ the Lorentzian 
configurations\footnote{ Even if one could accompany the Wick rotation 
with various other analytical continuations of the parameters 
describing the solution ($e.g.$ ${\hat a}\to i {\hat a}$ for a MPAdS$_5$ black hole), 
given the numerical nature of the 
 configurations in this paper, there is no  assurance that the modified metric functions 
will also be solutions of the field equations in Euclidean signature.
Instead, one has to solve directly the EGB field equations for a metric ansatz with Euclidean signature.}. 
In view of this difficulty one  has to resort to an alternative, 
quasi-Euclidean approach as described in \cite{quasi}.
The idea is to regard the action $I$ used in the 
computation of the partition function as a functional 
over complex metrics that are obtained from the real, 
stationary, Lorentzian metrics by using a transformation 
that mimics the effect of the Wick rotation $t\ra i\tau$. 
 In this approach, the values of 
the extensive variables of the complex 
  metric that extremize the path integral are the same as the values 
 of these variables corresponding to the initial Lorentzian metric\footnote{Note 
 also that not all closed form solutions with
Lorentzian signature present reasonable Euclidean counterparts, in which case
one is forced again to consider a 'quasi-Euclidean' approach.
The $d=5$ asymptotically flat rotating black ring solutions
provides an interesting example in this sense \cite{Astefanesei:2005ad}.}. 

 When computing the classical bulk action evaluated on the equations of motion,
one replaces the $R-2\Lambda+\frac{\alpha}{4}L_{GB}$ volume term with
$2(R_t^t+\frac{\alpha}{4}H_t^t)$ and make use of (\ref{totder1}) to express
it as a difference of two boundary integrals.
A straightforward calculation using the asymptotic
expansion (\ref{inf1})  shows that the divergencies of the boundary integral  at infinity,  together with the contributions
from $I_{b}^{(E)}$ and $I_{b}^{(GB)}$, are regularized by $I_{\mathrm{ct}}$. 
As a result, by using also the first set of relations in (\ref{totder1}),
 one finds the finite expression of the classical action
\begin{eqnarray}
\label{Icl} 
I_{cl}=\frac{V_3}{4  G}\bigg(
\frac{1}{\sqrt{f_1 b_1}}
\bigg[
(f_2-4b_2)\sqrt{1 - \frac{2\,\alpha}{\ell^2}}
+\frac{3\ell^2}{8}(1-\frac{6\,\alpha}{\ell^2} +\sqrt{1 - \frac{2\,\alpha}{\ell^2}}~)
\bigg]
\\
\nonumber
-\sqrt{h_h}(r_h^2+\alpha(4-\frac{4h_h}{r_h^2}))
-\frac{4}{\sqrt{f_1 b_1}}w_hw_4\sqrt{1 - \frac{2\,\alpha}{\ell^2}}
\bigg).
\end{eqnarray}
Upon application of the Gibbs-Duhem relation to the partition 
function,  one finds  the entropy 
$S=\beta (E- 2\omega_h J)-I_{cl}$, which is the sum of one quarter of the event horizon area plus a
Gauss-Bonnet correction
\begin{eqnarray}
S=S_0+S_{GB},~~{\rm with}~~S_0=\frac{V_3}{4G}r_h^2\sqrt{h_h},~~
S_{GB}=\alpha\frac{ V_3}{4G}\sqrt{h_h}(4-\frac{h_h}{r_h^2}).
\end{eqnarray}
In the static limit, the known expression
 $S = \frac{V_{3}}{4 G} (r_h^3 + 3 \alpha r_h)$
is recovered,
while the entropy of the MPAdS$_5$ solutions is $S= {V_3}A_H/{4G}$.

\section{Numerical results}
The EGB equations (\ref{eqs}) lead  to a system of four coupled second order ODEs 
for the metric functions\footnote{These equations
are extremely complicated (each of them containing around  fifty terms) and we shall not present them here.} $f(r),~b(r),~h(r)$ and $w(r)$.
We are interested in solutions of these equations presenting the asymptotic expansion (\ref{c1}), (\ref{inf1}).

\newpage
\setlength{\unitlength}{1cm}
\begin{picture}(18,7)
\centering
\put(2,0.0){\epsfig{file=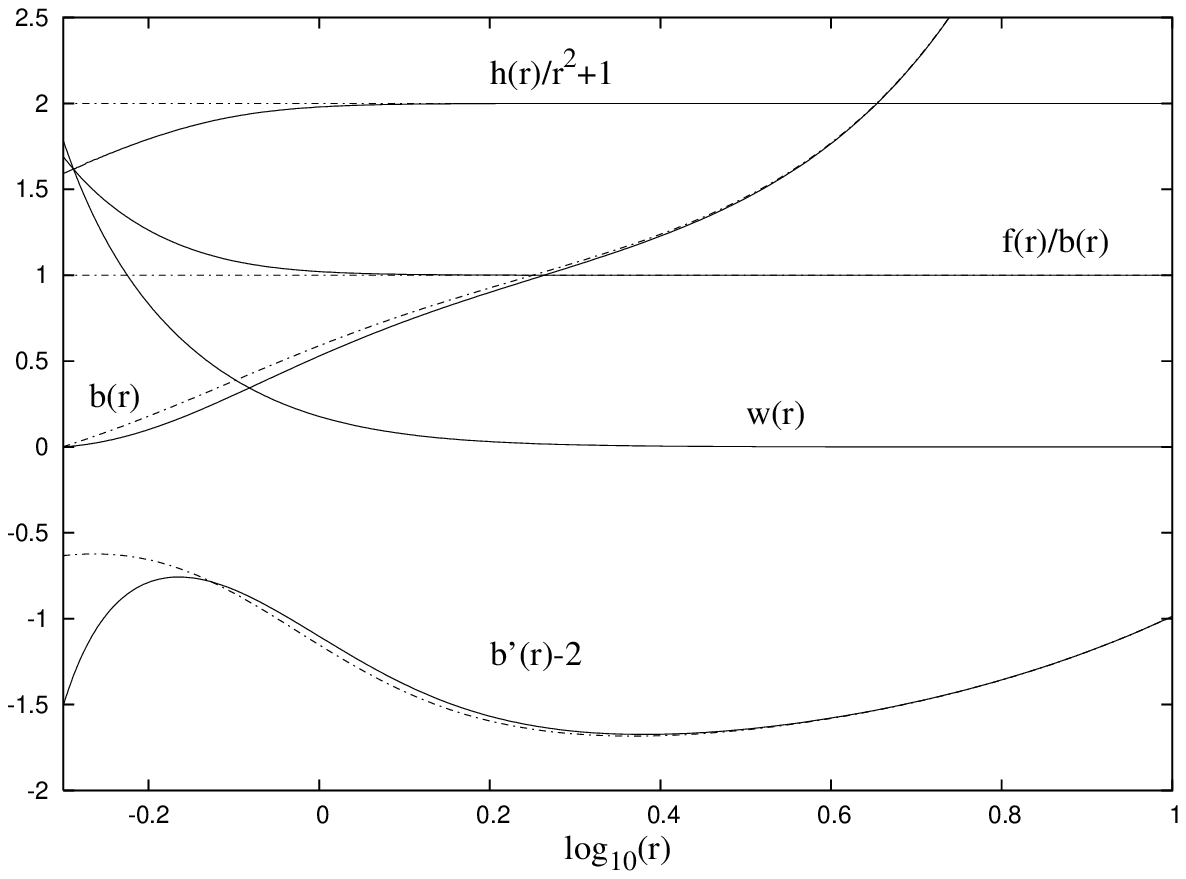,width=12.7cm}}
\end{picture} 
\\
\\
{\small {\bf Figure 1.} A typical rotating solution with $w_h=1.8$ (solid line) is plotted together
with a static solution ($w_h=0$, dashed line) for $r_h=0.5,~\alpha=0.5,~\ell^2=20$. 
 }
\\
\\
\\
\setlength{\unitlength}{1cm}
\begin{picture}(19,9)
\centering
\put(2.7,.0){\epsfig{file=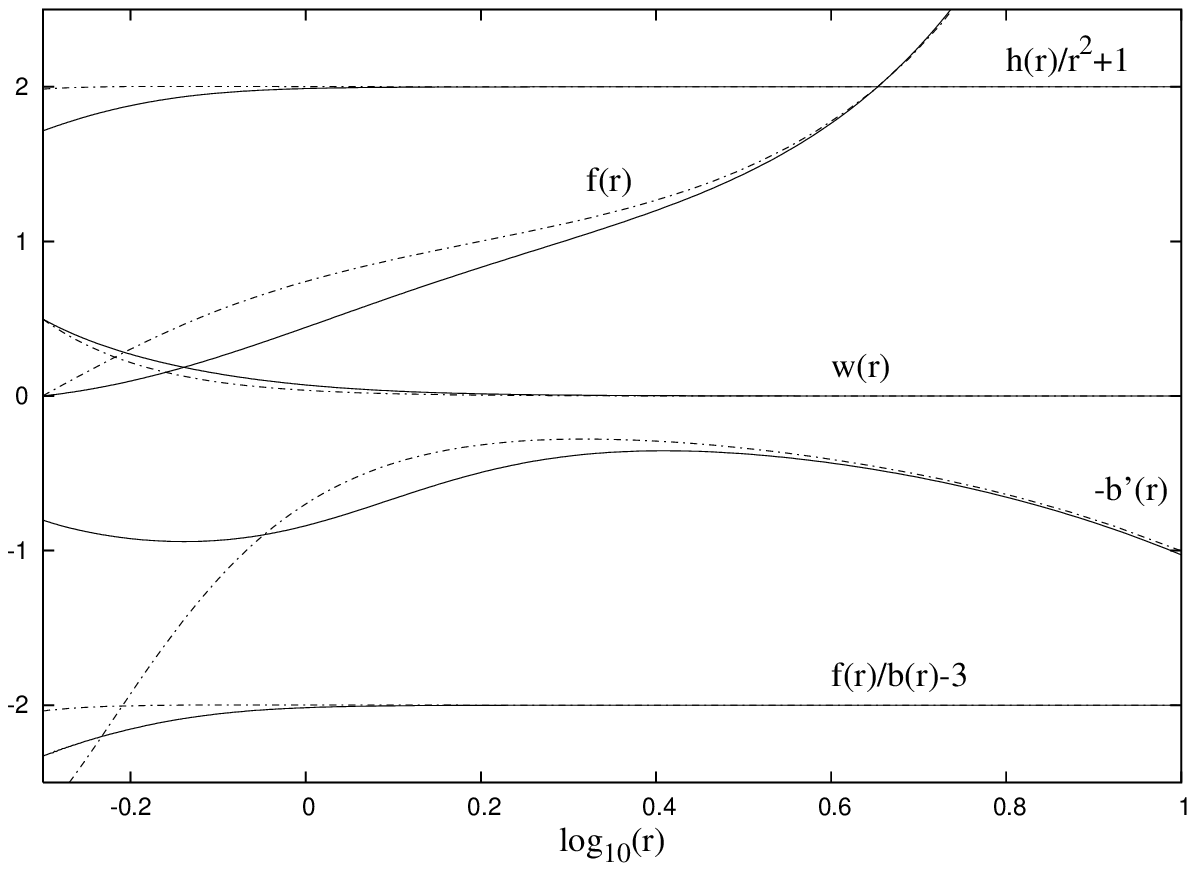,width=12.5cm}}
\end{picture} 
\\
\\
{\small {\bf Figure 2.}
Two rotating solutions with  $r_h=0.5,~w_h=0.5,~\ell^2=20$
are plotted for $\alpha=0.1$ (dashed line) and  $\alpha=1$ (solid line).}
\\
\\
In order to construct numerical solutions, the  constants $(\alpha,\Lambda)$ 
have to be
fixed. Then the solution is further specified by the event horizon $r_h$ and the 
angular velocity at the horizon $w(r_h)$
(or equivalently, the angular momentum
$J$ through the parameter $w_4$).

The complete classification of the solutions in the space 
of parameters is a considerable task that is 
not aimed in this paper. Instead, by taking the arbitrary value $\ell^2=20$,
we analyzed in detail a few particular 
\newpage
\setlength{\unitlength}{1cm}
\begin{picture}(18,7)
\centering
\put(2,0.0){\epsfig{file=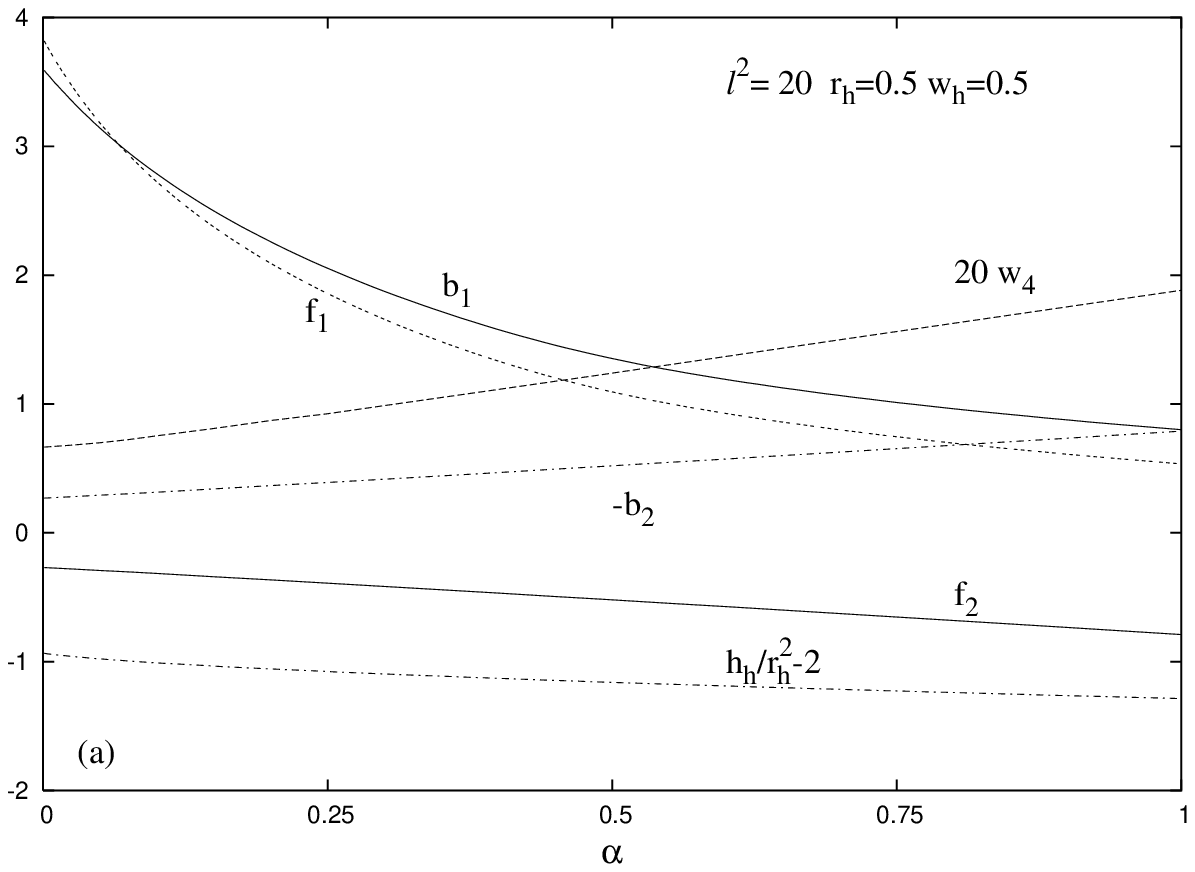,width=12.75cm}}
\end{picture}
\begin{picture}(19,9.4)
\centering
\put(2.2,0.0){\epsfig{file=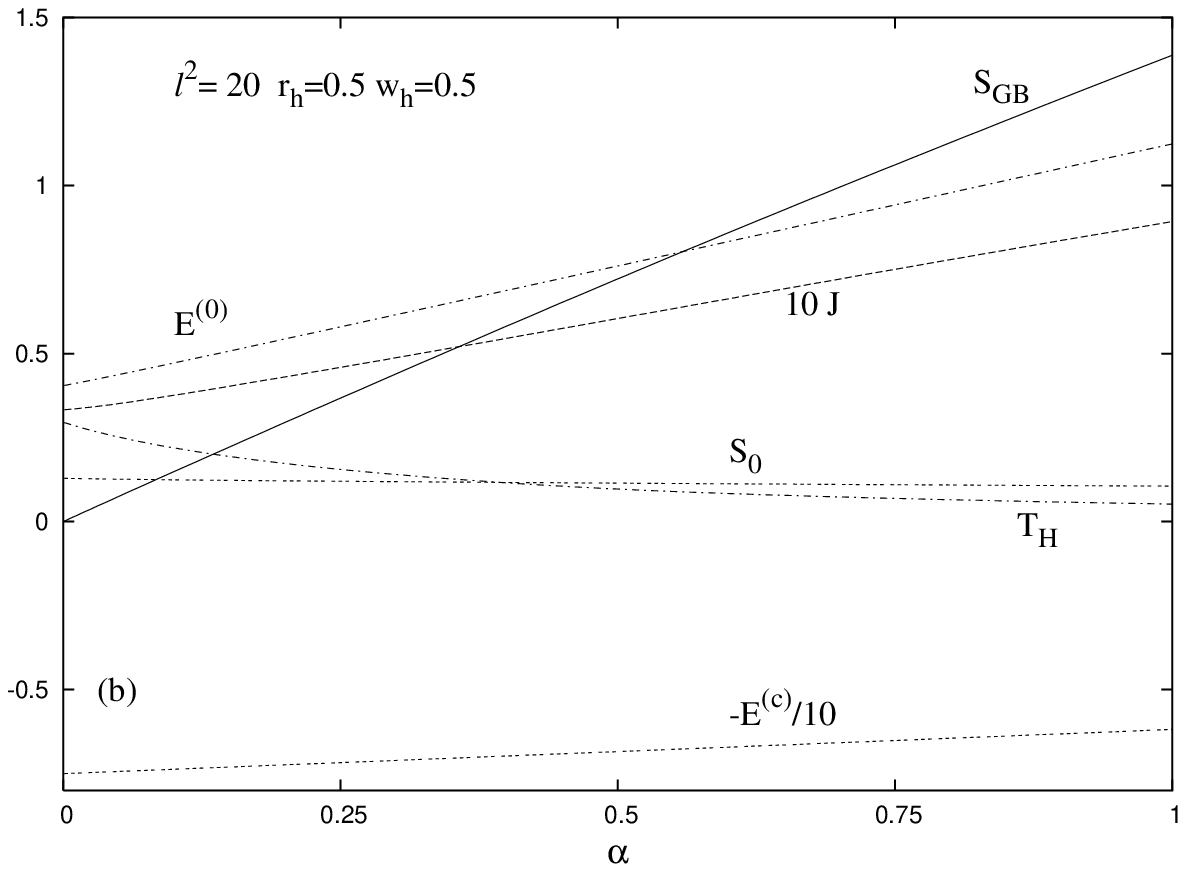,width=13cm}}
\end{picture}
\\
\\
{\small {\bf Figure 3.}
The parameters  $f_1,b_1,h_h$ at the event horizon 
are plotted together with the parameters  $f_2,b_2,w_4$
in the asymptotic expansion at infinity, as a function of $\alpha$
for solutions with $r_h=0.5,~\ell^2=20,~w_h=0.5$ (Figure 3a).
In Figure 3b we plot the Hawking temperature, the  mass-energy $E^{(0)}$, the Casimir term $E^{(c)}$, the 
angular momentum $J$ and the entropies $S_0,S_{GB}$ for these solutions.  Here and in Figure 4b $E^{(0)}$, $E^{(c)}$  and $J$
are plotted in units with $V_3/8\pi G=1$, while we set $V_3/4 G=1$ in the expression of $S$.}
\\
\\
classes of solutions, 
which hopefully would reflect all relevant properties 
of the general pattern.
However, we have found nontrivial rotating black hole solutions for other values of the cosmological constant,
in particular for $\Lambda=0$ and for $\Lambda > 0$.

Also, since the Gauss-Bonnet term in (\ref{action}) has to be considered as a correction
to the Einstein-Hilbert action, we report here the results for
positive values of  $\alpha$ in the interval $0 \leq \alpha \leq 1$ (however,
solutions with $\ell^2=20$ and larger values of $\alpha$
exist as well).

\newpage
\setlength{\unitlength}{1cm}
\begin{picture}(18,7)
\centering
\put(2,0.0){\epsfig{file=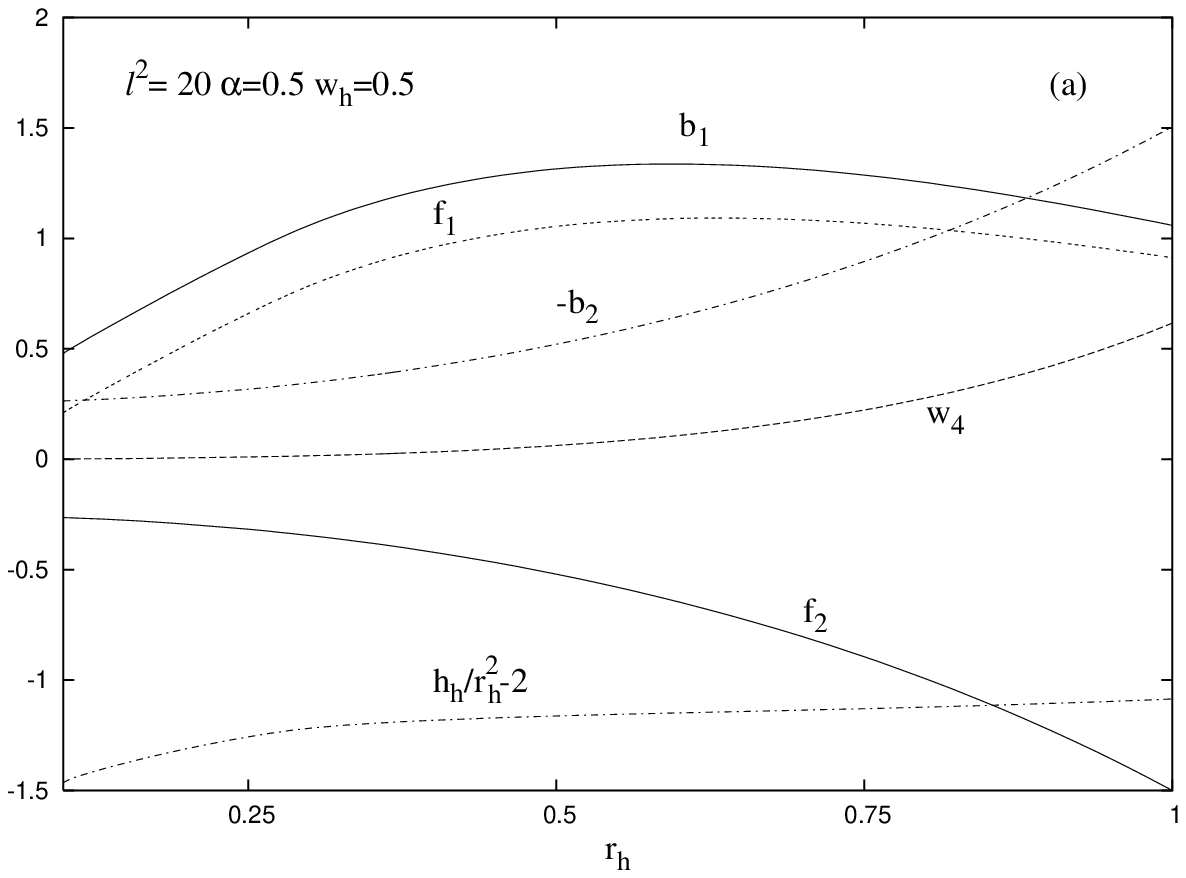,width=12.7cm}}
\end{picture}
\begin{picture}(19,9.4)
\centering
\put(2.5,0.0){\epsfig{file=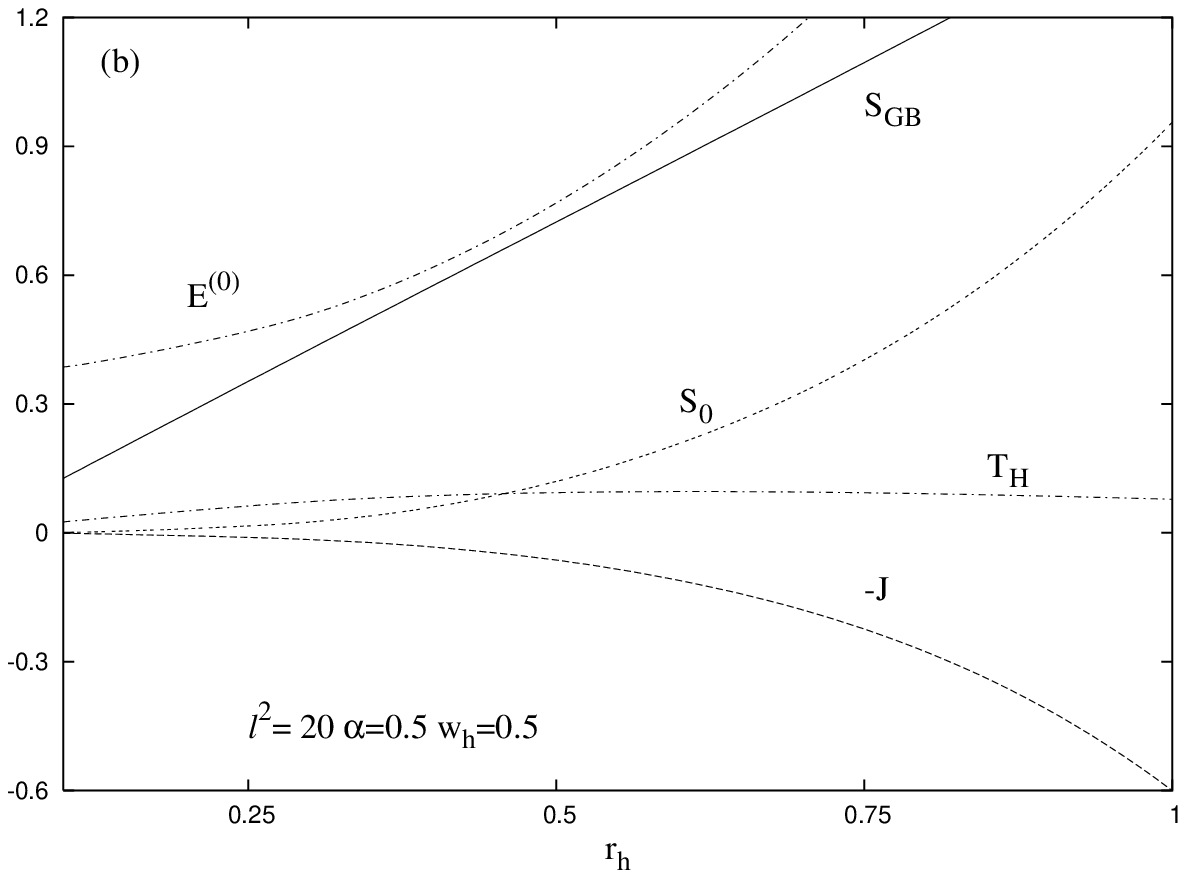,width=13cm}}
\end{picture}
\\
\\
{\small {\bf Figure 4.}
The parameters  $f_1,b_1,h_h$ at the event horizon are plotted together with the parameters  $f_2,b_2,w_4$
in the asymptotic expansion at infinity, as a function of the event horizon radius $r_h$
for solutions with $\alpha=0.5,~\ell^2=20,~w_h=0.5$ (Figure 4a).
In Figure 4b we plot the Hawking temperature, the  mass-energy $E^{(0)}$,  the 
angular momentum $J$, the entropy of the solution in Einstein gravity $S_0$ and the Gauss-Bonnet correction $S_{GB}$ for these solutions.}
\\
\\
In the absence of a closed form solution,
we relied on a numerical methods to solve the equations. 
The 
numerical methods here are similar to those  used in literature to find 
other numerical black hole solutions with equal
magnitude angular momenta  \cite{Kunz:2005nm,Kunz:2006eh}. 
We take units such that $G=1$,
and employ a collocation
 method for 
boundary-value ordinary
differential equations, equipped with an adaptive mesh selection procedure
\cite{COLSYS}.
Typical mesh sizes include $10^3-10^4$ points.
The solutions have a typical relative accuracy of $10^{-8}$. 

In constructing rotating EGB-AdS black holes, 
we make use of the existence of the closed form solutions
(\ref{SGB-AdS}) and (\ref{MP-AdS}), and employ them as starting configurations,
increasing  gradually $w_h$ or $\alpha$, respectively.

The profiles of the metric functions of a typical EGB-AdS black hole solution
corresponding to $\alpha = 0.5$, $r_h=0.5$
are presented on Figure 1 
for a static ($w_h=0$) and a rotating solution with $w_h=1.8$. 
One can see that the rotation leads to non constant values for $h(r)/r^2$ and $b(r)/f(r)$,
and is  particularly apparent on the function $b(r)$ and its derivative.
 
It is also natural to study how the profile of a rotating solution ($e.g.$ with a given angular velocity $w_h$)
is affected by the Gauss-Bonnet term.
This is  illustrated on Figure 2 where 
the profiles corresponding to $\alpha=0.1$ (dashed curves) --very close to the MPAdS$_5$ solution,
and $\alpha = 1$ (solid lines) are superposed for $r_h=0.5,~w_h = 0.5$.
 One can see there that the $r^2-$terms 
start dominating  the profile of the metric functions $f,b,h$ very rapidly, which implies 
a small difference between different solutions for large enough $r$.
However, the situation is different in the small$-r$ regime (see also Figure 1).  

 
We also performed an analysis of the EGB-AdS solutions 
when  varying the Gauss-Bonnet coupling 
constant $\alpha$. 
In the limit $\alpha=0$, the MPAdS$_5$ black holes (\ref{MP-AdS}) are recovered\footnote{With the 
particular values $r_h=0.5,~w_h=0.5$ that we have chosen to perform the numerical analysis,
the parameters $\hat a,\hat M$ of the MPAdS$_5$ solution correspond to $\hat a=10/81$ and $\hat M=6561/48640$.}.
The evolution of the parameters $ f'(r_h),b'(r_h),h(r_h)$ and $f_2,b_2,w_4$ characterizing the solutions 
is shown  on Figure 3a as function of $\alpha$.  
The corresponding physical quantities, as computed according to the relations in the previous Section,
are reported on Figure 3b.
 
 We also varied the event horizon radius $r_h$ for a set of given
$\alpha,~w_h$ and found
no evidence of a maximal value of $r_h$  where the solutions could eventually
terminate. 
The evolution of the solution data as a  function of the event horizon radius is reported on Figure 4a,
the mass-energy, angular momentum, entropy and Hawking temperature
 being plotted in Figure 4b.
For small values of $r_h$ the numerical analysis is quite
tedious and it strongly suggests that the derivatives of $w(r)$ and $h(r)/r^2$ become infinite
in the $r_h \to 0$ limit.   

Finally, although the numerics is more involved in this case,
we constructed solutions with fixed $\alpha$ and $r_h$
but varying the horizon velocity $w_h$. 
Equivalently, this leads to a family of solution with varying the angular 
momentum $J \sim w_4$
since there is a one-one correspondence between $w_4$ and $w_h$.
Similar to the $\alpha=0$ case, for each set of solutions we 
observe two branches, extending up to a maximal value of $w_h$, 
where they merge and end. The lower
branch emerges from the static solution in the limit  $w_h=0$.  
The maximal value of $w_h$ depends on the horizon
radius $r_h$,   the cosmological constant $\Lambda$, and the coupling constant $\alpha$.

\section{Further remarks }
The main purpose of this paper was to
present arguments for the existence of 
rotating black holes in $d=4+1$ EGB theory with negative cosmological
constant.
These configurations posses
a regular horizon of spherical topology and 
have two  equal-magnitude angular momenta, representing generalizations
of a particular class of MPAdS$_5$ black holes.
We also proposed to adapt the boundary counterterm formalism of 
\cite{Balasubramanian:1999re} to $d=4+1$ EGB-$\Lambda$ theory, computing in this way
the mass-energy and angular momenta of  solutions.
The general relations in Section 2.1 apply also to other known solutions in EGB theory with negative cosmological
constant and can easily be generalised for a positive sign of $\Lambda$.

The  solutions in this paper may provide a fertile ground for further study of
 rotating configurations in EGB theory.
For example, their generalization to include the
effects of  an electromagnetic field is straightforward.
Also,  in principle, by 
using  the same techniques,
 there should be no difficulty to construct similar solutions in $d=2N+1$ dimensions
 with $N>2$ equal magnitude angular momenta.
 An interesting problem here is to find the boundary counterterm expression 
 in EGB$-\Lambda$ theory for other values of $d>5$.
 
In the five dimensional case, one can also approach  the general case of a black hole with two
distinct angular momenta, by solving a set of partial differential equations with a
dependence on $(r,\theta)$.
The formalism proposed in Section 2 to compute the mass, angular momentum and entropy of AdS solutions
should apply in the general case, too.
Rotating topological black holes  in EGB theory 
with an horizon of negative curvature  are also likely to exist for $\Lambda<0$.

The study of the solutions discussed in this paper in an AdS/CFT context is an
interesting open question.
According to the
AdS/CFT correspondence, the higher derivatives curvature terms can be viewed as the corrections of
large $N$  expansion of the boundary CFT in the strong coupling limit.
For the ansatz considered here, the boundary metric
is not rotating and corresponds to a static Einstein universe in four dimensions.
Here it is interesting to note that, similar to the $\alpha=0$ case, the stress-energy tensor  
for the dual theory defined in that background,  as computed according to the standard prescription \cite{Myers:1999qn},
is traceless. 

A detailed study of the $d=4+1$ rotating black hole solutions in EGB theory 
together with a discussion of their asymptotically 
flat limit will be presented elsewhere.
\\
\\
{\bf\large Acknowledgements} 
\\
Y. B. thanks  the
Belgian FNRS for financial support.
\\
\\ 
{\it  Note added:} When this work has been at final stages, 
the authors of \cite{Alexeyev:2007sd} appear to have succeeded in constructing 
a Kerr black hole in EGB theory in $d=4+1$ dimensions. 
However, this result needs an independant confirmation.
Afterwards, it would be interesting to extend
 the results in Section III of  \cite{Alexeyev:2007sd}
to the case of black holes with negative cosmological constant and 
two equal magnitude angular momenta.

 \begin{small}
 
 \end{small}

\end{document}